\def\no{\noindent}

\def\\{$\backslash$}
\def\deg      {{\ifmmode^\circ\else$^\circ$\fi} } 
\def\arcm     {{\ifmmode {'\ }\else$'     $\fi} } 
\def\arcs     {{\ifmmode{''\ }\else$''    $\fi} } 

\def\arcspt   {{$\buildrel{\prime\prime}\over .$}}

\def\cle      {{$_ <\atop{^\sim}$}}

\def\Lsun     {{$L_{\odot}$} }

\documentstyle[12pt,aasms4,psfig]{article}

\begin{document}

\noindent {\it Dissertation Summary} 

\begin{center}
\title{\large \bf The Nature of Radio Emission from Distant Galaxies} 

	E. A. Richards, Department of Physics \& Astronomy \\
              Arizona State University, Box 871504 \\
             Tempe, AZ 85287-1504; eric.richards\@asu.edu \\     

\vspace{5pt}

\noindent Thesis Work Conducted at: National Radio Astronomy Observatory 
               \& University of  
  Virginia 

\noindent Ph. D. thesis directed by K. I. Kellermann; degree awarded August 1999

\end{center} 


        In this thesis I present the observational
results of a multi-wavelength campaign aimed
at investigating the nature of radio emission
from distant galaxies, and in particular to understand their
 implication for star-formation at early epochs.
The radio observations
were centered on the Hubble Deep Field (HDF; Williams, R. 1996, AJ, 112, 1335)
and conducted at 8.5 GHz and 1.4 GHz using 
the Very Larg Array (VLA) and the Multi-Element Microwave 
Linked Interferometer (MERLIN). These observations represent
the most sensitive radio survey thus far, achieving
rms flux densities of 1.6 $\mu$Jy (8.5 GHz) and 4 $\mu$Jy 
(1.4 GHz) respectively, at resolutions from 
0\arcspt 2 to 6\arcs (Richards, E. A. et al. 1998, AJ, 116, 1039;
Richards, E. A. 1999, ApJ, in press; Muxlow, T. W. 1999 in preparation). 
The results of the radio study and optical identifications of the 72 radio
sources detected in a complete sample ($S_{1.4} \geq$ 40 $\mu$Jy)
located within the HDF and its flanking fields show that:     

\begin{itemize}

\item  The 1.4 GHz direct source count remains steep to 
       40 $\mu$Jy
       with a differential slope of 2.4. 
      Fluctuation analysis suggests continuity
      in the count to $\sim$1 $\mu$Jy, implying
      a surface density of discrete radio sources
      of 60 arcmin$^{-2}$. There is evidence
      for field to field count variations and
      spatial clustering within the 40\arcm VLA field 
       of view.

\item	The mean spectral index of the 8.5 GHz selected
      sample is $\overline{\alpha } _{8.5}$ = 0.35 ,
      in good agreement with previous studies (Fomalont, E. B. et 
       al. 1991, AJ, 102, 1258; Windhorst, R. A. et al. 1993, 
       ApJ, 405, 498). Only 
      20\% of the 8.5 GHz selected sample can be
      attributed directly to AGN activity and we 
      place an upper limit of 15\% on the fraction 
      of inverted sources selected at 8.5 GHz. Several
      of the radio sources are identified
      with star-forming galaxies and their flat spectrum radio
      emission ( $0 < \alpha < 0.5$; $S_{\nu } \propto
      \nu ^{- \alpha}$) may be associated with an enhanced 
      fraction of Bremsstrahlung radiation.
      The 1.4 GHz selected sample has a mean spectral
      index of $\overline{\alpha } _{8.5}$ = 0.85 ,
      somewhat higher than that of shallower samples. 

\item Based on 151 sources in the complete
      VLA 1.4 GHz sample, the average angular size 
      is observed to decrease as a function of flux density.
      At $S_{1.4}$ = 370 $\mu$Jy, 
    $\theta _{med}$ = 2.6\arcsec~$\pm$ 0.4\arcsec , 
    and at $S_1.4$ = 100 $\mu$Jy , $\theta _{med}$ = 
    1.6\arcsec $\pm$ 0.3\arcsec .  This is in good
    agreement with the complimentary MERLIN 
    observations (Muxlow, T. W. et al. 1999, in preparation) of 91 of the above 
    VLA sources.
    For those MERLIN sources with $S_{1.4} >$ 90 $\mu$Jy
    the mean is 1.7\arcs $\pm$0.3\arcs   
    and for those sources with $S_{1.4} <$ 90 $\mu$Jy,
    the mean was determined to be 1.2\arcs $\pm$0.1\arcs .

\item Eighty percent (64) of the 79 radio sources in the
    HDF region were identified in the HST I-band 
    (F814W) and deep ground-based images (Barger, A. J.
    et al. 1999, AJ, 117, 102) to $I_{AB}$ = 25, with a median of
    $I_{AB}$ = 22.1. From the 72 radio sources
    contained within the HST frames for which
    morphological classification was possible, 80$\pm$10\% are associated
    with disk galaxies (spirals, irregulars, peculiars, mergers),
    with a lesser fraction (20$\pm$10\%) identified with bright
    ellipticals.

\item Twenty percent of microjansky radio sources cannot
  be identified to $I_{AB}$ = 25 in the HST frames
  and to a similar depth in ground-based images.
  Three sources remain unidentified in the HDF
  to $I_{AB}$ = 28.5.

\end{itemize}

	Based on radio spectral index, radio morphology, and
  galaxy type of the optical identification, we classified the
  origin of the radio emission for individual detections 
 (Richards, E. A. et al. 1998).
  For 60\% of the identifications with $I_{AB} <$ 25, the 
  radio emission was characterized as principally star-forming.
  Twenty percent were classified as AGN, and the remaining sources
  were ambiguous. Thus the bulk of the microjansky 
  radio population is associated with star-forming galaxies
  in the redshift range 0.2 $< z <$ 1.3 (Barger, A. J.,  Cowie, L. L \& Richards,
   E. A. 1999, AJ, submitted).      

	The nature of the optically faint ($I_{AB} >25$) radio 
  sources remains unclear. Three of these source contained
  in the HDF are likely associated with independently 
  detected millimeter sources at 0.85 mm and 1.3 mm 
  (Richards, E. A. 1999, ApJL, 513, 9; Hughes, D. L. et al. 1998, Nature,
  394, 241; Downes, D. et al. 1999, A\& A, 347, 809),
  suggesting interpretation as high redshift dust enshrouded 
  starbursts. Based on the far-infrared to radio flux density relationship
  observed in local starburst galaxies (e.g., Condon et al. 1991, 
   ApJ, 378, 65),
  these galaxies are likely at 1 \cle $z$ \cle 3 and have 
  far-infrared luminosities of 10$^{12-13}$\Lsun (Barger, A. J. Cowie, L. L.
  \& Richards, E. A. 1999, AJ, submitted). 
  However, one radio source (VLA J123646+621226, detected independently
  at 8.5 GHz and 1.4 GHz) contained in the Thompson/NICMOS
  deep field was detected neither in the mm
  nor in NICMOS J/H to 29th magnitude (Thompson, R. I. et al.
  1999, AJ, 117, 17).  

	From the HDF radio survey and another with similar
 radio sensitivity in Selected Area 13 (Windhorst et al.
 1995; Richards et al. 1999), I have isolated a sample
 of 32 optically faint and/or very red ($I-K \geq$ 4-6)
 objects, comprising 20\% of the microjansky counts
 to $S \simeq$ 10 $\mu$Jy with a surface density of
 $\sim$0.2 arcmin$^{-2}$. These sources constitute
 a new class of radio objects which are likely a 
 heterogeneous mixture of 1) dust-enshrouded 
 starburst galaxies at 1 \cle $z$ \cle 3 , 2)
 dust reddened AGN at moderate redshift (1\cle $z$ 2\cle ;
 Dunlop et al.,1996, Nature, 381, 581), and extreme redshift 
($z >$6) nascent AGN 
 embedded within proto-ellipticals
 in the process of initial collapse (Richards, E. A. et al. 1999,
 ApJL, in press).
 
	Figure 2 shows a
 radio image for a proposed confusion limited
 deep survey with the VLA in its most extended
 A configuration (rms noise $\sim$ 0.12 $\mu$Jy
 with 2\arcs resolution) 
 with a detection limit of about 1$\mu$Jy.  
 At these flux levels the radio sky may become 
 increasingly 
 dominated by the optically 'invisible' population. 
 Future radio surveys with
 increased resolution and sensitivity, together with 
 complimentary sub-mm imaging made possible by 
 planned facilities such as the Expanded VLA,
 the Square Kilometer Array, and the Atacama Large
 Millimeter Array will be necessary to characterize
 this new radio population and further discern its
 nature.

\newpage

{\noindent \bf Figure 1:} Montage of radio/HST flanking field overlays. 
Contours are 1.4 GHz fluxes drawn at 2, 4, 8, 16,
 32, 64 $\sigma$ ($\sigma$ = 4 $\mu$Jy). Greyscale
is log stretch of HST I-band image 5\arcs
on a side. {\bf VLA J123725+621128:} One of only
two classical radio sources found in our HDF radio
survey. This wide-angle tail galaxy is identified
with a $K = 20.3$ elliptical. {\bf VLA J123634+621212:}
$I_{AB}$ = 21.1 disk galaxy is at $z = 0.45$
with a steep radio spectral index ($\alpha$ = 0.7).
HST image shows evidence of a double nucleus and
a dust lane coincident with the radio source.
70\% of radio selected field galaxies fit into   
this starburst category.
{\bf VLA J123634+621240:} A dramatic
$z = 1.3$ starburst, with
3 $\times$ the luminosity of Arp 220.
The peculiar optical ID has $I_{AB}$ = 23.5.
{\bf VLA J123642+621331:}
An unidentified source with $I >$ 25, subsequently identified
with an H = 22 galaxy and with a $z = 4.42$
(Waddington, I. et al. 1999, ApJL, submitted).
Some 20\% of the $\mu$Jy radio sources remain optically
 unidentified.                                      

{\noindent \bf Figure 2:} Shown is a simulation of a      
confusion limited HDF sized region at 1.4 GHz
as seen by the proposed Expanded VLA.
The true field of view is 200$\times$ larger,
with a total of 40,000 detections above
0.8 $\mu$Jy (100$\times$ deeper than the
current VLA/HDF survey). Such observations
could detect the Milky Way at $z = 1$ and
Arp 220 up to $z = 10$ free from dust
obscuration. 

\slugcomment{{\it PASP Dissertation Summary}}

\newpage

\begin{center}
{\large \bf Acknowledgments}
\end{center}

	My sis -- thanks for
always encouraging me to 
do well and pep talks.

	My bro randy -- thanks for
the help observing. It was fun.

	NRAO - for providing me with the
tools and support to do this research.

	Bob O. -- for not calling
the cops during my North Carolina
hurricane search and rescue adventure. Also
for the Kitt Peak observing trip
and making me \$ 2 richer! 

	Tom Muxlow -- for honing my skills as a
radio astronomer and being a great host
at Jodrell (even if I had to live in a flooded
apartment). Also letting me swipe some
of his impressive figures for this thesis.

	Anita Richards -- thanks for taking care
of me and making even Macellsfield an interesting
place. The Dog was fun.

	Rogier Windhorst -- for putting
me up during the summer of 1996 in the
Tempe roach motel.
Tempe. And also for being a great collaborator.
Can't wait to use those Fiesta Bowl tickets
next year.

	Bruce Partridge -- for being a great
teacher at and beyond Haverford and first 
introducing me to the art of radio astronomy.
I look forward to many years of future collaboration.
	
	Ed Fomalont -- for showing me 
the tricks of the trade. Also for being
patient with my antics.

	Ken Kellermann -- for being the bestest
thesis advisor a student could ask for. I learned
a lot more than astronomy from you.

Amy Barger and Len Cowie -- for providing optical and
 near infrared photometry. Also for starting
 what has already proved to be an extremely fruitful
 collaboration.

	Rowan-Robinson -- for making research
entertaining.

	Jody, Jason, Curt, and all my friends
back home -- thanks for the good times and providing
much needed diversions.

Susan --  I have never seen a woman ride
a horse bare-back until I met you. The
summer of 1998 was one of the best of my
life, take care.

	Mom and Dad --- thanks for giving
me the heart and the means to pursue 
my career.

	Mike P. -- thanks for being a 
patient roomate. And for providing 
countless evenings of grad student
angst venting.

	Steve Balbus -- for having faith
in me -- it meant so much.

The General Lee (a.k.a., the red Porsche) --
I have never had one car get me in so
much trouble - from speeding tickets to
claims of racism. It was worth it! I
hope your new owner takes better
care of you than I.

	Ronak -- thanks for putting me
up at your place, providing rides, 
and being a friend. You the man, nak!

	Siegel -- how could I have made it
without another Southern boy around with
reasonable politics. Sorry about the Braves.    

	Chris -- all I can say is Tennessee
rocks. Thanks for hosting all the football
and grad parties up at Alden house.

	Ray -- Thanks for
just being a great class mate
the first two years of school, 
it was definitely a dual effort --
I still can't
believe we both passed the qualifier.
Also for reminding me that I wasnt
the grad student of southern descent.

	Majewski -- for providing 
sage advice at critical times.

	Tolbert -- for introducing me
to the mysteries of Campari.

	Hibbard -- for helping me 
jump off my bike. Also the good 
career advice was key. I wont forget
you when I become famous.

Jennifer -- for adding flavor to the
department. Good to know that the 
South is still alive and well in
the Forestry building.

Franz -- Hang in there bud. It's almost
Miller time. See you in Hawaii on Y2K.

\end{document}